# Extension, Torque and Supercoiling in Single, Stretched and Twisted DNA Molecules

Pui-Man Lam[*]
Physics Department, Southern University
Baton Rouge, Louisiana 70813

and

Yi Zhen[+]
Department of Natural Sciences, Southern University
6400 Press Drive
New Orleans, Louisiana 70126

**Abstract:** We reinvestigate the model originally studied by Neukirch and Marko that describes the extension, torque and supercoiling in single, stretched and twisted DNA molecules, which consists of a mixture of extended and supercoiled state, using now a more accurate form of the free energy for the untwisted but stretched DNA. The original model uses an approximate form of this free energy and the agreement with experiment is only qualitative. We find that this more accurate free energy significantly improves the results, bring them into quantitative agreement with experiment, throughout the entire force regime. This is rather surprising, considering that the theory is completely parameter-free.

## I. Introduction

Cooperative topological transition of a DNA molecule is of great importance in a number of essential cellular processes involving rotation or torque generation. Strand separation of the double helix is required in initiation of both replication and transcription, facilitating the accessibility of its bases to a variety of protein. Alteration of the topological state of DNA plays important regulatory roles in transcription. Torque is generated in transcription and it potentially affects the structure of chromatin. Replication is coupled with unwinding of the parental DNA strands and buildup of torsional stress [1-6]. Thus it is crucial to elucidate the mechanisms of behavior of DNA under torque and force.

DNA molecules with no nicks (no break in one of the strands) have twist modulus, thanks to their double-helix structure. This property has important biological consequences. First this provides an efficient way to compact DNA to fit into cells and nuclei. Secondly, a twisted DNA molecule can be either more stable or less stable than its untwisted form depending on whether the twist is negative (underwind) or positive (overwind), respectively. Thus a variety of proteins (RNA polymerase [7], regulation factors [8], etc.) can locally underwind and hence denature the bases of DNA, making them more easily accessible. On the other hand, thermophilic bacteria that live in environment close to the boiling point of water have enzymes (reverse gyrases [9]) that overwind the molecule and hence making it stable even under such conditions. Because topology plays such an important role in cell life, Nature has evolved a family of enzymes, generally known as topoisomerases [10-16] (reverse gyrase being one of them) that control the torsion and entanglement of the molecules.

It is now well known that DNA molecules can be stretched and twisted using single manipulation experiments [10, 15-22], following the seminal work of [22]. In these experiments, a DNA molecule is anchored at multiple points to a surface and to a bead used to apply a force and a torque and the extensions of the molecule as a function of double helix linking number are measured. Mosconi et al.[21] used a magnetic trap system to measure the buckling torque of DNA as a function of force in various salt concentration. Results of experiment showed that supercoiled DNA has a mixed-phase regime where extended and plectonemically supercoiled phases coexist when buckling torque of the molecule is above the threshold.

The semi-flexible polymer model provides a quantitative starting point for theories describing these type of experiments [23-30]. In ref. [27] Marko has suggested a heuristic model to describe the behavior of a stretched DNA under twist. In this model, the twisted, stretched molecule is partitioned into an unstretched, plectonemic supercoil phase with torsional stiff $P$ and a stretched and twisted DNA molecule with bending rigidity $A$ and effective torsional stiffness $C_s$. The torsional stiffness $P$ of the plectonemic DNA is unknown and can only be estimated. The predictions of the model are in qualitative agreement with experiment [21]. This is not surprising since a description of the plectonemic phase with a single force independent torsional stiff $P$ is an oversimplification that does not take into account for example the variation of plectonemic radius with force due to entropic repulsion [31].

More recently, Neukirch and Marko [28] proposed an improved model that describes the stretched, twisted DNA molecule by taking into account the variation of the plectonemic radius with force explicitly. In this way the theory is completely parameter-free. In the limit of high force, closed-form asymptotic solutions for the supercoiling radius, extension and torque of the molecule can be obtained. These asymptotic solutions already give rather reasonable agreement with experiment. In this paper we want to show that much of the discrepancy of this theory with experiment is due to the use of an approximate form of the free energy for the extended phase of the molecule, which is stretched but untwisted. Use of a more accurate form of the free energy significantly improves the agreement with experiment.

In section II we will briefly recapitulate the Neukirch-Marko model. In section III we will show our calculation with an improved form of the free energy. Section IV is the conclusion.

## II. The Neukirch-Marko Model

We will follow ref. [28] in order to introduce their model. Fig. 1 is an illustration of a supercoiled DNA molecule under force $f$ and torque. The total length $L$ of the molecule is partitioned between the two "phases": (i) a plectonemic phase of length $l$, where the filament has bending rigidity $A$ and torsional rigidity $C$ and adopts a superhelical shape of radius $r$ and angle $a$, and (ii) an extended wormlike-chain phase of length $L-l$.

The free energy of the extended phase is described in terms of the free energy per unit length of the untwisted molecule $g(f) = f - k_B T \sqrt{f/A}$ [26], plus a twist energy using a twist modulus that includes effects of writhing fluctuations: $C_s(f) = C\left[1-(C/4A)k_B T/\sqrt{Af}\right]$ [32]. Both forms are correct for large $f$. For simplicity, both in ref. [28] and in this paper, the form $C_s = C$ will be used. But we will show in the next section that it is the use of the above approximate form for $g(f)$ that has led to much of the discrepancy with experiment.

The free energy of the plectonemic phase is that of two straight charged cylinders with a center axis separated by a distance of $2r$, in the Debye–Hückel approximation of the Poisson-Boltzmann equation. For the double helix, two negative charges appear for each base pair. This suggests the use of a linear charge density (in electron charge units) of $n = 1/b$, where $b = 0.17$ nm is half of the 0.34 nm spacing of successive base pairs along DNA. However, an effective charge is introduced to cope with two effects: (i) the fact that this charge is distributed on the surface of the cylindrical double helix of radius $a = 1$ nm rather than on its center axis and (ii) the asymptotes of the linear and nonlinear solutions of the Poisson-Boltzmann equation have to match for large separation distances. The effective charge is

$$n = \frac{1}{b} \frac{1}{g(L_B b, k_D a)} \frac{1}{k_D a K_1(k_D a)} \qquad (1)$$

where $L_B = 0.7$ nm is the Bjerrum length in water, $k_D^{-1}$ the Debye length, and $K_n(x)$ the nth modified Bessel function of the second kind [33, 34]. From Table III of ref [33], the parameter $g$ is computed to be $g = (1.64, 1.44, 1.27, 1.14)$ at salt concentrations (50, 100, 200, 500) mM and for $T = 296.5$ K.

The interaction potential in the plectoneme is [26, 35]

$$U(r) = k_B T n^2 L_B K_0(2 k_D r) \qquad (2)$$

where both $k_D$ and $n$ depend on the salt concentration.

By adding together electrostatic, bending, and twisting energy terms, the total free energy is

$$F(a, r, l_s, t_p, l) = -g(L - l) + \frac{1}{2} C l_s^2 (L - l)$$
$$+ \left[ \frac{1}{2} C t_p^2 + \frac{1}{2} \frac{\sin^4 a}{r^2} + U(r) \right] l \qquad (3)$$

where $l_s = 2p \Delta Lk_s / (L - l)$ is the linking angle density in the stretched part of the DNA ($\Delta Lk_s$ is the excess linking number of the extended region) and $t_p = 2p \Delta Tw_p / l$ is the twist angle density in the plectonemic DNA ($\Delta Tw_p$ is the excess twist in the plectoneme region). In the second term we have used $C_s(f) = C$ for simplicity.

Once force and $\Delta Lk$ are specified, the remaining variables are determined by the minimization of Eq. (3), subject to the constrainst

$$\Delta Lk = \Delta Tw + Wr = \frac{1}{2p}\left(\mathbf{l}_s(L-l) + \mathbf{t}_p l + \frac{\sin 2\mathbf{a}}{2r}l\right) \quad (4)$$

where $\Delta Lk$ is the number of turns introduced into the DNA relative to the relaxed double helix linking number $Lk_0$ (i.e., $\Delta Lk = Lk - Lk_0$).

The constraint on $\Delta Lk$ is handled by using a Lagrange multiplier $M$, i.e. by minimizing $G = F - 2pM(\Delta Tw + Wr - \Delta Lk)$. Using Eqs. (3) and (4), this becomes

$$G = -g(L-l) + \frac{1}{2}C\mathbf{l}_s^2(L-l) + \left[\frac{1}{2}C\mathbf{t}_p^2 + \frac{1}{2}\frac{\sin^4 \mathbf{a}}{r^2} + U(r)\right]l$$
$$- M\left(\mathbf{l}_s(L-l) + \mathbf{t}_p l + \frac{\sin 2\mathbf{a}}{2r}l\right) \quad (5)$$

Equilibrium values of the six variables $\mathbf{l}_s$, $\mathbf{t}_p$, $M$, $\mathbf{a}$, $l$, $r$ follow from setting to zero the partial derivatives of $G$ with respect to these six variables. The partial derivatives with respect to the first three variables result in

$$C\mathbf{l}_s = C\mathbf{t}_p = M \quad , \quad \text{or} \quad \mathbf{l}_s = \mathbf{t}_p = M/C \quad (6)$$

$$\mathbf{l}_s(L-l) + \mathbf{t}_p l + \frac{\sin 2\mathbf{a}}{2r}l = 0 \quad (6a)$$

Using Eq. (6) this becomes

$$l = L\frac{2\mathbf{l}_s r}{\sin 2\mathbf{a}} \quad (7)$$

The partial derivatives with respective to the last three variables yield three other

equations:

$$-A\frac{\sin^4 a}{r^3}+U'(r)+M\frac{\sin 2a}{2r^2}=0 \qquad (8)$$

$$\frac{2A}{r}\sin^3 a \cos a - M\cos 2a = 0 \qquad (9)$$

$$g(f)+\frac{1}{2}A\frac{\sin^4 a}{r^2}+U(r)-\frac{\sin 2a}{2r}M=0 \qquad (10)$$

The three equations (8), (9) and (10) can be solved to obtain the remaining three unknowns quantities $a$, $r$ and $M$.

In the high-force limit, Neukirch and Marko [28] have given the solutions in closed form. Define a quantity $K$ as

$$K=\sqrt{9p/8n^2 L_B k_B T/g(f)} \; . \qquad (11)$$

The plectoneme radius $r$ and angle $a$ are given by

$$r = \log K/(2k_D) \qquad (12)$$

$$a = \left[2r^2 g(f)/(3A)\right]^{1/4} \qquad (13)$$

The torque $M$ is given by

$$M = \frac{2A}{r}\frac{\sin^3 a \cos a}{\cos 2a} \qquad (14)$$

A Taylor expansion of this expression for small $a$ and substituting the results, Eqs. (12) and (13) for $a$ and $r$ yields

$$M \approx [(32/27)A]^{1/4} g(f)^{3/4} / \sqrt{k_D} \sqrt{\log\left[\sqrt{9\boldsymbol{p}/8}\boldsymbol{n}^2 L_B k_B T / g(f)\right]}(1+\boldsymbol{a}^2) \qquad (15)$$

Another quantity of experimental interest is the slope of the average extension $q = \partial <X> / \partial \Delta Lk = -\partial^2 G / \partial f \partial \Delta Lk = -2\boldsymbol{p}\partial M / \partial f$. Using Eq. (10) for $M$, this becomes

$q = -4\boldsymbol{p} r g'(r) / \sin 2\boldsymbol{a}$. Taylor expanding this for small $\boldsymbol{a}$ and substituting Eqs. (12) and (13) for $\boldsymbol{a}$ and $r$ yields

$$q = \left(\frac{6A}{k_D^2 g(f)}\right)^{1/4} \sqrt{\log\left[\sqrt{9\boldsymbol{p}/8}\boldsymbol{n}^2 L_B k_B T / g(f)\right]} g'(f)(1 + 2\boldsymbol{a}^2/3) \qquad (16)$$

Using Eqs. (15) and (16) and $g(f) = f - k_B T \sqrt{f/A}$, the slope and torque calculated are in qualitative agreement with experiment [28]. Experimental values of $A/k_B T = 46$, 47, 44, 45 nm at 50, 100, 200, and 500 nM salt and $C/k_B T = 94$ nm are used in the calculation. We will show in the next section that using a more accurate form of $g(f)$ can significantly improve on the agreement with experiment.

**III. Calculation using an improved free energy**

In this section we give our calculation of the slope and the torque using an improved form of the untwisted free energy.

The force-extension curve in the worm like chain (WLC) model is given by the widely used interpolation formula [26]

$$f = \frac{(k_B T)}{L_p}\left[\frac{X}{L} + \frac{1}{4}\left(1 - \frac{X}{L}\right)^{-2} - \frac{1}{4}\right], \qquad (17)$$

where $L_p$ here is the persistence length, related to the bending rigidity $A$, by $A = k_B T L_p$,

and $X$ is the extension. The negative of the free energy per unit length $g(f)$ is obtained by a Legendre transform

$$Lg(f) = fX - W(X) \qquad (18)$$

where

$$W(X) = \int_0^X dX' f(X') \qquad (19)$$

is the work done in extending the polymer. The functions $g$ and $W$ depend also on the persistence length $L_p$. From Eqn. (17) the extension $X$ is an implicit function of the force $f$. Since the extension is a single-valued, monotonic increasing function of $f$, we can define the inverse function $X_f(f)$ which give the extension $X$ as a function of the force $f$. Even though this function cannot be obtained analytically, it can be calculated numerically to high accuracy. Substituting Eqn. (17) into Eqn. (19), the function $W$ can be calculated analytically:

$$W(X(f)) = L\frac{k_B T}{4L_p}\left[\frac{X_f(f)}{L}\left(2\frac{X_f(f)}{L}-1\right)+\left(1-\frac{X_f(f)}{L}\right)^{-1}\right] \qquad (20)$$

From Eq. (18), the negative of the free energy per unit length is given as a function of the force $f$ by

$$g(f) = \frac{1}{L}fX_f(f) - \frac{k_B T}{4L_p}\left[\frac{X_f(f)}{L}\left(2\frac{X_f(f)}{L}-1\right)+\left(1-\frac{X_f(f)}{L}\right)^{-1}\right]$$
(21)

We will use this form of the free energy in Eqs. (15) and (16) to calculate the torque

$M$ and slope $q$.

From Eq. (16), in order to calculate the slope $q$, the derivative of $g$ with respect to $f$ is needed. From Eqs. (18) and (19), this is given by

$$Lg'(f) = X_f(f) \qquad (22)$$

In Fig. 2 we show our calculation of the slope of the average extension $q = \partial <X>/\partial \Delta Lk$ obtained using Eqn.(16), with $g$ and $g'$ given by Eqns (21) and (22), together with results obtained using the approximate forms for $g$ and $g'$. The experimental data are directly taken from Fig. 2 of ref. [28]. The data in ref. [28] are obtained from ref. [21]. The slope in ref. [21] is a dimensionless quantity defined as $\tilde{q} = L^{-1}\partial <X>/\partial s$, with $s = (Lk - Lk_0)/Lk_0$, where $Lk_0 \approx 1500$ is the linking number of the DNA molecule under no external tension or torque. Our slope $q$ is related to $\tilde{q}$ by $q = (L/Lk_0)\tilde{q} = (5.4\text{mm}/1500)\tilde{q} = (3.6nm)\tilde{q}$. The experimental data given in ref. [28] is actually a factor $(-3.6nm)$ times the data in ref. [21]. In Figure 3 we present the results of our calculation for the torque $M$, using Eqns. (15) and (21), together with results obtained using the approximate form, compared with the experimental data, taken from Fig.3 of ref. [21]. We can see that this better form of the free energy improves significantly the agreement with experiment. The agreement with experiment is now surprisingly good, except for low salt concentrations.

## IV. Conclusion

We have shown that by using a better form of the free energy for the stretched but untwisted part of the DNA, the Neukirch-Marko model can give quantitative agreement with experimental results. There are still some disagreement at low salt concentration, but this is probably due to the inadequacy of the Debye–Hückel approximation of the Poisson-Boltzmann equation, which results in imperfect screening of the electrostatic potential at these low salt concentrations. It was mentioned in ref. [28] that the disagreement with experiment may be due to the neglect of confinement entropy [36]. Since our results using a better free energy already yield quantitative agreement with experiment, the effect of confinement entropy is probably small.

Our calculation is based on the model of Neukirch and Marko [28]. This theory is an analytic theory, with analytic expressions for the slope and torque as functions of the tension. In order to arrive at this theory, several reasonable simplifications have been introduced. It does not incorporate thermal fluctuations in plectoneme. The argument is that at least at higher tensions, the fluctuations are small and can as a consequence be neglected. It also neglects multi-plectoneme effects. The use of a two cylinder repulsion in the Debye–Hückel regime is a rough approach not taking into account the effect on plectoneme angle as was shown to be important by Ubbink and Odijk [35]. More recent models [37,38] have taken these effects into account. In ref. [37] the authors give results of the slope versus tension, in very good agreement with experiment. However, for this quantity, the original theory of Neukirch and Marko also gives good agreement with experiment. It is the torque versus tension results in the Neukirch-Marko theory that show the largest disagreement with experiment, especially for low tension and low salt concentrations. For the torque versus tension result, the result of ref[37] is not so good. In more recent work [38] Marko and Neukirch have also incorporated the above mentioned effects in their model, but unfortunately they do not give any new torque versus tension results.

Notwithstanding the clearly better agreement between theory and experiment achieved in this work, one notes, however, that it holds as far as the Debye–Hückel approximation of the Poisson-Boltzman theory remains valid, i.e., for high screening/salt concentration only. As one can see from Figs. 2 and 3, the agreement with experiment deteriorates at low salt concentration for both the slope and the torque.

A closer inspection of the q-f variation, shown in Fig. 2, indicates that the agreement with experiment at higher applied tensions (when f>3 pN at 500 mM and f>1 pN at 200 mM). This is puzzling because the expressions for the free energy $g(f)$ and twist modulus $C_s(f)$ should be correct for large $f$ and the fluctuations in plectoneme and multi-plectoneme effects neglected in the model, should also decrease with tension. Dhar et al [39] and Samuel et al [40] have explored effects that go beyond the high force limit ($g(f) = f - k_B T \sqrt{f/A}$). At these high forces, such effects may be relevant. It should also be pointed out that the force-extension formula given in Eqn. (17) is only an interpretation formula which is convenient for calculation and should not be considered as a substitute for analytical or semi-analytic theoretical models in ref. [39, 40]. In particular, the force-extension curve given by Eqn. (19) did not take into account the entropy of the chain, even when no external force is applied, as pointed out by Neumann [41]. Also, in this high tension limit one would have to include the effects of thermal fluctuations on DNA elasticity, as studied by Kulic et al [42,43] and Sinha et al [40,44]. Finally, the Legendre transform (Eqns. (18) and (19)) was used because we are considering the long DNA limit. If one were to look at shorter chains, with chain lengths comparable to the persistence length of 50 nm, one would need to work with Laplace transforms instead [41,45,46]. This is because such short chains are not in the thermodynamic limit and one has to distinguish between the isometric ensemble in which the chain ends are held fixed and the applied force is allowed to fluctuate and the isotensional ensemble in which the applied force is held fixed and the chain lengths are allowed to fluctuate. Only in the thermodynamic limit in which the chain lengths are infinitely long do the two ensembles yield identical results.

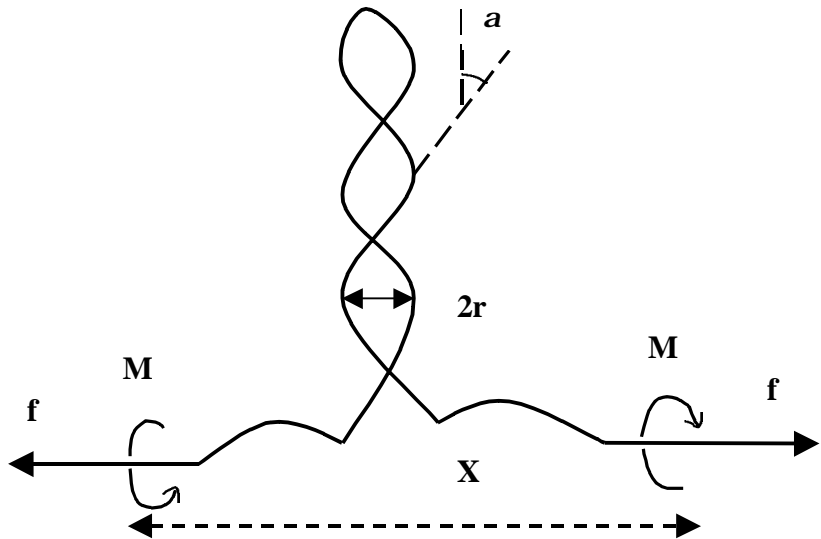

Figure 1. Supercoiled DNA under force and torque. Molecule length is partitioned between two phases: an extended phase and a plectonemic phase where strong self-interaction occurs.

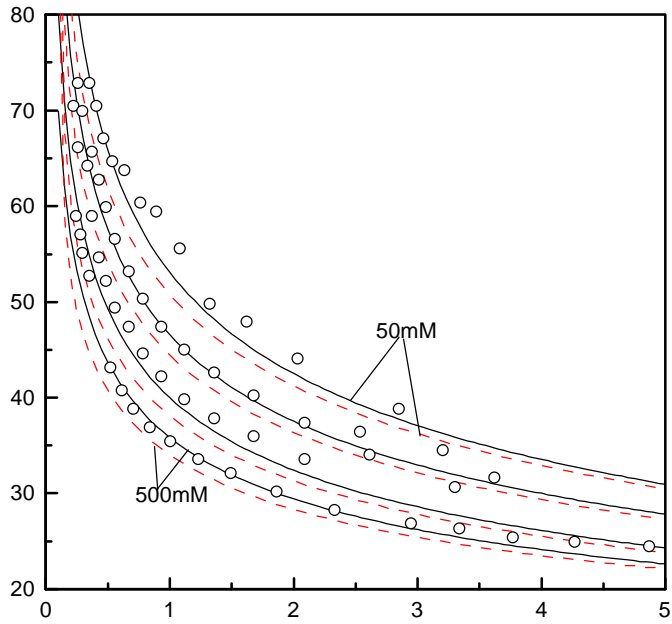

Fig. 2. Comparison of experimental and theoretical slopes $q = \partial <X> / \partial \Delta Lk$ of the average extension, as a function of the applied force, for 50, 100, 200 and 500 mM salt (top to bottom). Circles are experimental data. Full lines are our theoretical results using better form of the free energy. Dashed lines are theoretical results using approximate form of the free energy.

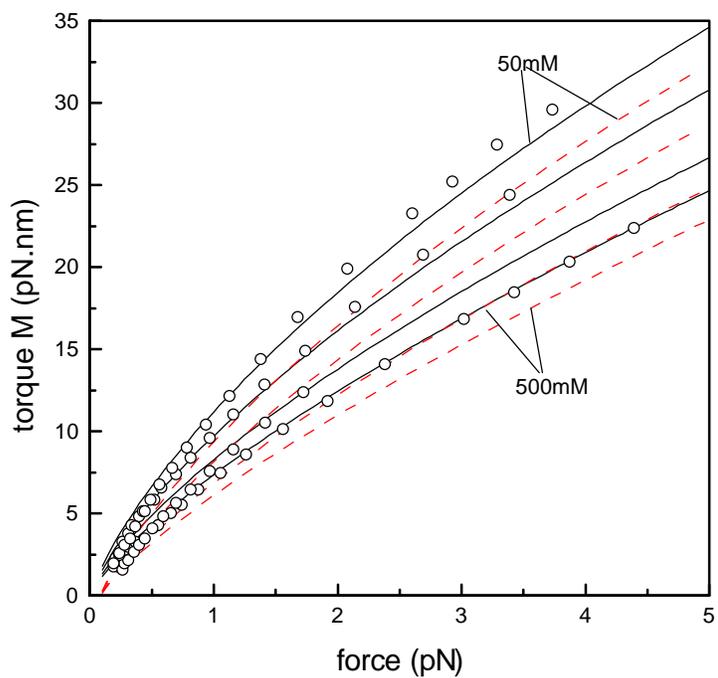

Fig. 3. Comparison of experimental and theoretical torque as a function of the applied force, for 50, 100, 200 and 500 mM salt (top to bottom). Circles are experimental data. Full lines are our theoretical results using better form of the free energy. Dashed lines are theoretical results using approximate form of the free energy.